\documentclass[onecolumn]{aastex631}

\usepackage{wrapfig}
\usepackage{rotating}

\newcommand\HST{{\it Hubble Space Telescope \/}}

\shorttitle{M87 Jet Novae}
\shortauthors{Lessing, Shara, Hounsell et al.}

\begin{document}


\title[M87 NUV novae]{A 9-Month {\it Hubble Space Telescope} Near-UV Survey of M87. II. A Strongly Enhanced Nova Rate near the Jet of M87}

\correspondingauthor{Michael Shara  mshara@amnh.org}

\author[0000-0001-6714-2706]{Alec M. Lessing}
\affiliation{Department of Physics, Stanford University, Stanford CA 94305, CA, USA}

\author[0000-0003-0155-2539]{Michael M. Shara}
\affiliation{Department of Astrophysics, American Museum of Natural History, New York, NY 10024, USA}

\author[0000-0002-0476-4206]{Rebekah Hounsell}
\affiliation{University of Maryland Baltimore County, 1000 Hilltop Cir, Baltimore, MD 21250, USA}
\affiliation{NASA Goddard Space Flight Center, Greenbelt, MD 20771, USA} 

\author[0000-0002-6126-7409]{Shifra Mandel}
\affil{Department of Astronomy, Columbia University, New York City, NY 10024, USA}

\author[0009-0008-7883-2129]{Nava Feder}
\affiliation{Yale University, New Haven, CT}

\author[0000-0002-9011-6829]{William Sparks}
\affil{SETI Institute, Mountain View, CA 94043, USA}

\begin{abstract}
The 135 classical novae that we have discovered in M87 with two {\it Hubble Space Telescope} imaging surveys appear to be strongly concentrated along that galaxy's jet. Detailed simulations show that the likelihood that this distribution occurred by chance is of order 0.3\%. The novae near the jet display outburst characteristics (peak luminosities, colors and decline rates) that are indistinguishable from novae far from the jet. We explore whether the remarkable nova distribution could be caused by the jet's irradiation of the hydrogen-rich donors in M87's cataclysmic binaries. This explanation, and others extant in the literature which rely on increased binary mass transfer rates, fail by orders of magnitude in explaining the enhanced nova rate near the jet. An alternate explanation is the presence of a genuine surplus of nova binary systems near the jet, perhaps due to jet-induced star formation. This explanation fails to explain the lack of nova enhancement along M87's counterjet. The enhanced rate of novae along M87's jet is now firmly established, and unexplained. 
\end{abstract}


\section{Introduction} \label{sec:intro}

Cataclysmic binaries comprise a white dwarf (WD) accreting matter from a hydrogen or helium-transferring companion star \citep{Warner1995}. The accumulation of $\sim 10^{-4} - 10^{-5}$ $M_{\odot}$ of hydrogen onto a WD \citep{starrfield1972,Shara1981,yaron2005} leads to a thermonuclear runaway that ejects much or all of the accreted envelope \citep{Menzel1933,Payne-Gaposhkin1977} at speeds of thousands of km/s \citep{slavin1995,santamaria2022}. This phenomenon is called a ``classical nova". The most luminous classical novae achieve luminosities L $\sim 10^{6}$ $L_{\odot}$, making them detectable as far away as the Virgo and Fornax galaxy clusters \citep{Pritchet1985,Shafter2000,Ferrarese2003,Neill2005,Curtin2015,Shara2016}. 

Large samples of extragalactic novae, all at the same distance, enable statistical studies of the distributions of luminosities, colors, and speed classes of these objects. Such studies have demonstrated important differences in the populations of bulge and disk novae in spiral galaxies \citep{Ciardullo1987,DellaValle1988,Shafter2001,darnley2006}. In contrast, the novae of the giant elliptical galaxy M87, which must have been accreted through multiple galaxy-cannibalization episodes, are thoroughly mixed \citep{Shara2016,Shara2023}. There is no correlation between an M87 nova's peak luminosity, color or rate of decline and its radial distance from the center of M87. However, the suggestion has been raised (based on a sample of just $\sim$ 13 novae \citet{Madrid2007}) that classical novae {\it do} seem to occur close to M87's jet at a rate higher than chance alone might dictate \citep{Livio2002}. No other galaxy with jets has been observed with sufficient sensitivity or frequency to yield samples of novae large enough to check if M87's putative nova-jet connection is ubiquitous, rare or spurious. 

We have recently found 94 erupting novae during a 5-day cadence, 260-days-long {\it Hubble Space Telescope} imaging survey of M87 \citep{Shara2023}, nearly tripling the number of novae known in that galaxy. This study confirmed that novae closely follow the K-band light of M87, and that the high nova rate first claimed by \citet{Shara2016}
is correct. Most novae in M87 (and by implication, those in other galaxies) have been missed by previous (ground-based) surveys because of sparse cadence, variable seeing and inability to detect the faint ``faint-fast" novae \citep{Kasliwal2011}. Our now much enlarged nova sample enables us to finally test the provocative suggestion that novae are distributed asymmetrically in M87, with a ``preference" for alignment with that galaxy's jet \citep{Madrid2007}. 

In Section \ref{sec:datasets} we describe the observational data that yielded the sample of 135 novae used in this paper's analysis. The definition of the M87 jet axis, the resulting angular distribution of the above 135 novae, and simple statistics and simulations that demonstrate ``clustering" of novae about the jet, are described in Section 3. Section 4 describes the simulations we carried out to quantify the effects of variable placement of the {\it HST} detectors, and the slightly non-spherical nature of M87. We use these simulations to quantify the enhancements of novae within different-shaped areas surrounding the jet in Section 5. In Section 6 we  compare the expected and real distributions of novae outside the region of the jet to see if any deviation of one from the other can be found. In Section 7 we contrast the novae near the jet with those further away, and we speculate on the cause of the nova rate enhancement near the jet in Section 8. We briefly summarize our results in Section 9.

\section{The Datasets}\label{sec:datasets}

Two deep \textit{Hubble Space Telescope} synoptic surveys with high cadence and long baselines have found a combined sample of 135 novae in M87. \cite{Shara2016} found 41 novae in 72 days of near-daily cadence HST ACS WFC observations, with most of the 61 epochs in the dataset having 500 seconds of integrated exposure time in $F606W$ and 1440 seconds in $F814W$. \cite{Shara2023} found 94 novae in 53 HST WFC3 epochs regularly spaced by 5 days and spanning 260 days, with 720 seconds of integrated $F606W$ exposure time and 1500 seconds of $F275W$ exposure time in most epochs.

{The 41 novae of \citet{Shara2016} were found with two independent searches. In the first, all F814W images of a given epoch were interlaced to produce a Nyquist image, which was then rectified and convolved with a filter designed to highlight point-source variables. Novae candidates were then identified in difference frames with a statistical criterion for variability above a baseline, with a final rejection of very faint candidates by visual inspection. In the second, Multidrizzle \citep{koekemoer2002hst} was used to created combined images of each epoch (and series of 5 subsequent epochs) were blinked and novae recovered via visual inspection.}

{The 94 novae of \citet{Shara2023} were also recovered with two independent searches. In the first, DizzlePac’s \citep{avila2014drizzlepac} \texttt{astrodrizzle} package was used to combine all FLC images of a given epoch and band. Nova candidates were found by visual inspection of and the use of SExtractor \citep{Bertin96} on difference images created from the drizzled images. In the second, lightcurves were created from drizzled images (made with FLCs from 1 epoch and sequences of 3 sequential epochs). Novae whose lightcurves showed statistically significant variability were flagged as candidates. The two lists of novae were combined and candidates with irregular or resolved PSFs, with too weak a signal to be confident in, or with lightcurves or colorcurves inconsistent with the the behavior of novae were removed.}

\begin{figure}[h]
    \centering
    \includegraphics[width=5.4in]{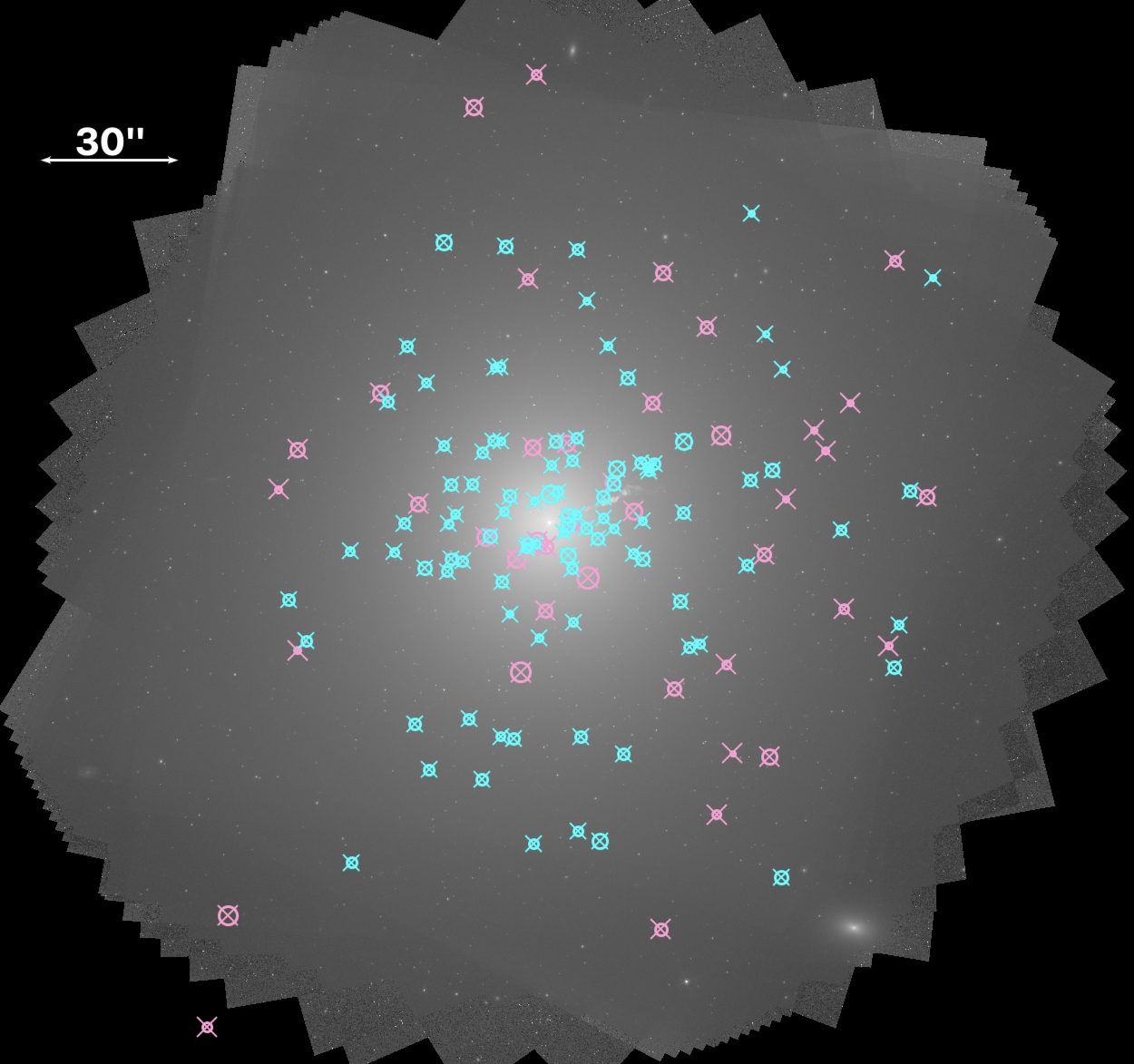}
    \vspace{0.2in}
    \includegraphics[width=5.4in]{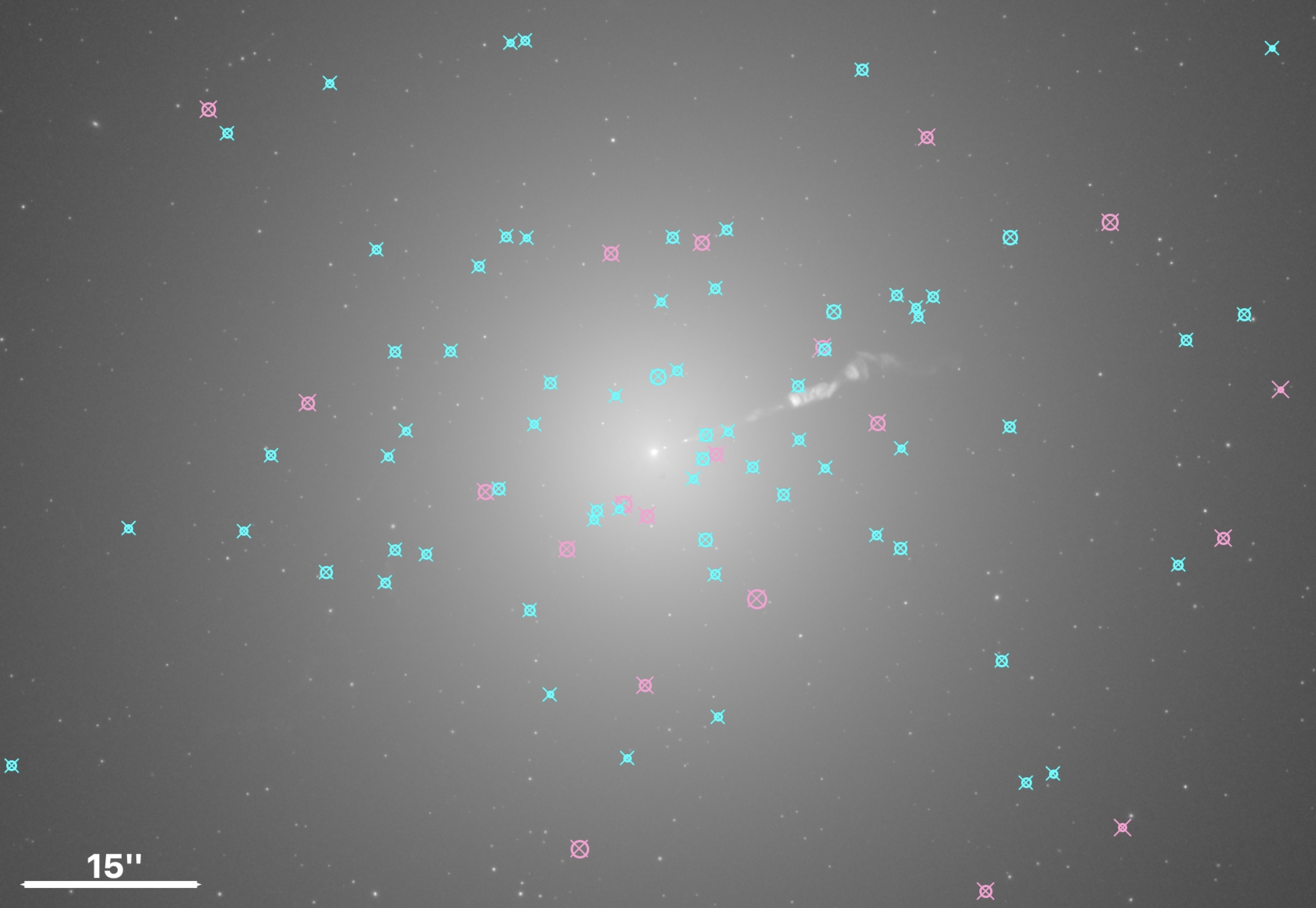}
    \caption{{Top: The Field of View of the 53 \HST pointings of the 2017 WFC3 dataset of \citet{Shara2023} and locations (cyan crosses) of all 94 novae detected in M87 in \citet{Shara2023}. Also shown (pink crosses) are the 41 novae of \citet{Shara2016}. North is up and East is left. The size of each nova's circle scales linearly with the brightest observed $F606W$ magnitude of that nova. Markers for novae whose peaks were not observed do not have a circle. Bottom: A close-up of the nuclear region of M87 and its novae.}}
    \label{fig:finderchart}
\end{figure}

Drizzled images of both datasets were aligned to Gaia stars \citep{Gaia2023} in the M87 field of view. The published positions of the 94 novae in \cite{Shara2023} were already aligned to Gaia stars. The 41 novae from \cite{Shara2016} were located by eye and new centroided positions were calculated in the Gaia-aligned frame; they have a $\sim$ 1.0" offset from the coordinates in \cite{Shara2016}. {These updated positions can be found in Table \ref{tab:updated-pos}. The positions of the novae from both datasets are plotted in Figure \ref{fig:finderchart}.}

\newpage

\section{The Nova Angular Distribution about the jet}\label{sec:region-enhancement}

\subsection{Locating the Jet}\label{ssec:locating-jet}

\begin{figure}[h]
    \centering
    \includegraphics[width=7in]{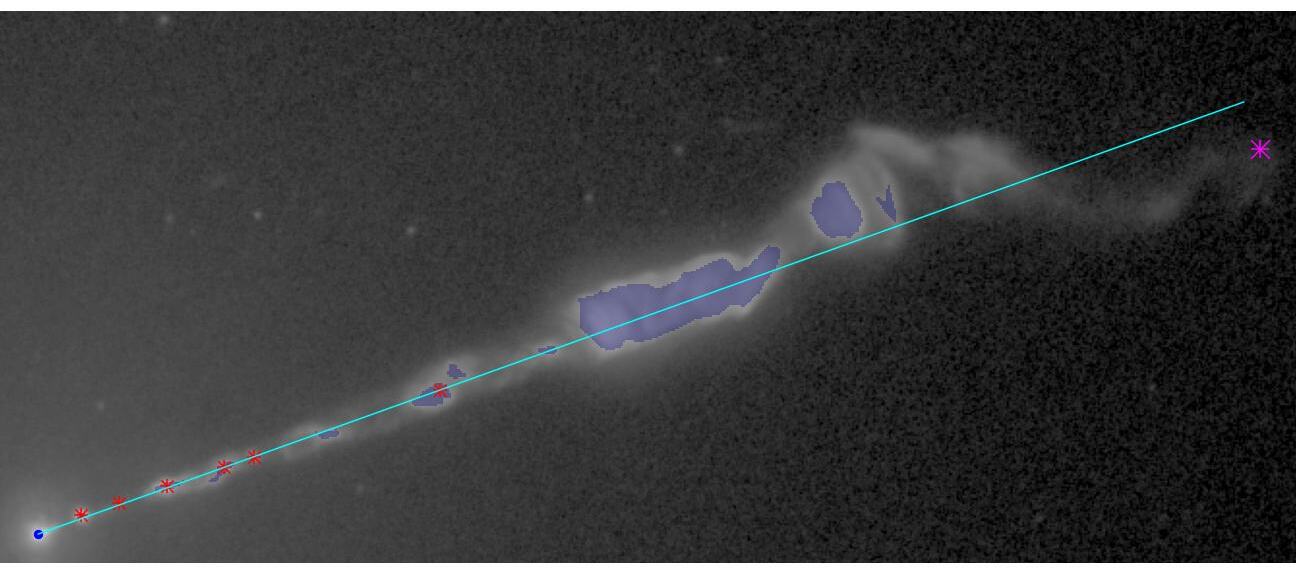}
    \caption{The jet centerline (cyan) used throughout this paper plotted on top of a drizzled image containing all $F275W$ data. The red points are the locations of the centroided point-like features used to fit the centerline to the image. Also shown (blue shading) are the region of the jet deemed bright enough that detecting a nova in that part of the image would be impossible with current data (see Section \ref{sec:sim-design}) and the furthest point on the jet deemed detectable (magenta).}
    \label{fig:jet-location}
\end{figure}

Several bright point-like sources in the jet can be used to unambiguously define the jet center-line. Their centroids were determined in a drizzled image made from all WFC3 $F275W$ data (the galaxy background was dimmest in $F275W$, allowing the jet to be analyzed with the greatest precision). A line segment extending from the center of M87 was fit to these centroided points such that the mean square distance of these points to the line segment was minimized. The resulting line was used as the jet center-line (see Figure \ref{fig:jet-location}). A radial coordinate system, used throughout this paper, was constructed centered at M87's center with angles measured counterclockwise relative to M87's jet center-line. The furthest point visually identifiable as part of the jet was measured at 26.2" from the center of M87, and this distance was adopted as the length of the jet.

\subsection{The novae}\label{ssec:the-novae}
\begin{figure}[h]
    \includegraphics[width=3.5in]{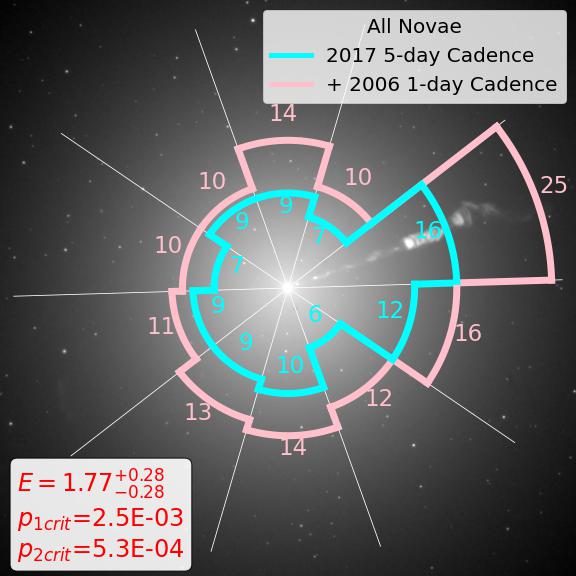}
    \includegraphics[width=3.5in]{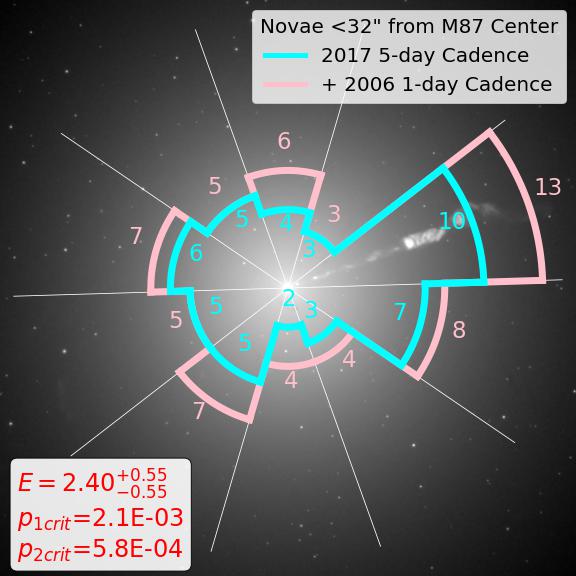}
    \caption{Histograms of the angular positions of all 135 novae (left) and of the 62 novae within 31.5" (120\% of the length of the jet) of M87's center (right). In each histogram, one of the histogram bins is centered on the jet, as described in section \ref{ssec:locating-jet}. $E$ (the nova rate enhancement) is the ratio of the number of novae within the jet-aligned histogram bin to the number expected if the nova distribution ``followed the K-band light" (see section \ref{sec:sim-rel-enhancement}). $p_{1crit}$ is the probability that the observed nova rate enhancement (relative to the expected, simulated distribution), or a higher enhancement, would occur randomly. $p_{2crit}$ is the probability that this would occur \textit{and} the maximum enhancement of any one of the other histogram bins would be no greater than the value actually observed. Different bin sizes yield different p values. See text for further details.}
    \label{fig:ten-wedges}
\end{figure}

In both datasets described in Section 2, the detector was centered close to M87's center (within $\sim 10$"). The high degree of radial symmetry of M87 ensures that variations of nova detection completeness with angle are small (though not zero). Figure \ref{fig:ten-wedges} shows histograms of the angular locations of the M87 novae. It is evident from this figure that the nova distribution within M87 is preferentially skewed in the direction of the jet. It is equally evident that there is no such enhancement in the direction of the counterjet \citep{Sparks1992}.

In the ten histogram wedges of Figure \ref{fig:ten-wedges}, 25 novae fall within the bin aligned with jet and no more than 16 fall in any other bin. The probability of seeing 25 or more novae in the jet-aligned bin if all 135 novae were distributed randomly among the ten bins is one in 541, as computed with the cumulative mass function of a $Binom(135, 0.1)$ binomial distribution. The probability of seeing at least 25 novae in that wedge {\it and} no more than 16 in any other wedge is one in 1310. To obtain this number, we performed 1,000,000 trials where we randomly assigned 135 novae to 10 bins and counted the number of trials in which both criteria were satisfied. Similar analysis on the 62 novae that lie within 32" (120\% the length of the jet) of M87's center shows 13 novae within the jet-centered wedge and no more than 8 in any other wedge, 
and yields probabilities of one in 115 and one in 345, respectively. These p-values -- computed assuming the simplest possible assumption, that the detected novae had an equal chance of falling in each wedge -- provide important ``sanity checks" of statistical significance.

\newpage

\section{Simulating the ``Expected" Distribution}\label{sec:sim-design}

 How good is the assumption of an expected uniform distribution of novae in angle about M87? In order to more precisely assess the observed nova distribution, an expected distribution of detected novae was computed for both datasets. These computations assumed that novae ``follow the K-band light," as found by previous studies (see Figure 3 of \cite{Curtin2015}, Figure 3 of \cite{Shara2023}, and Section \ref{sec:distr-away-from-jet} of this paper). Additionally, they account for effects on the nova distribution from the slightly variable positioning of the detector in each epoch and the slightly variable local nova detection efficiency of each dataset.

\begin{figure}[h]
    \centering
    \plottwo{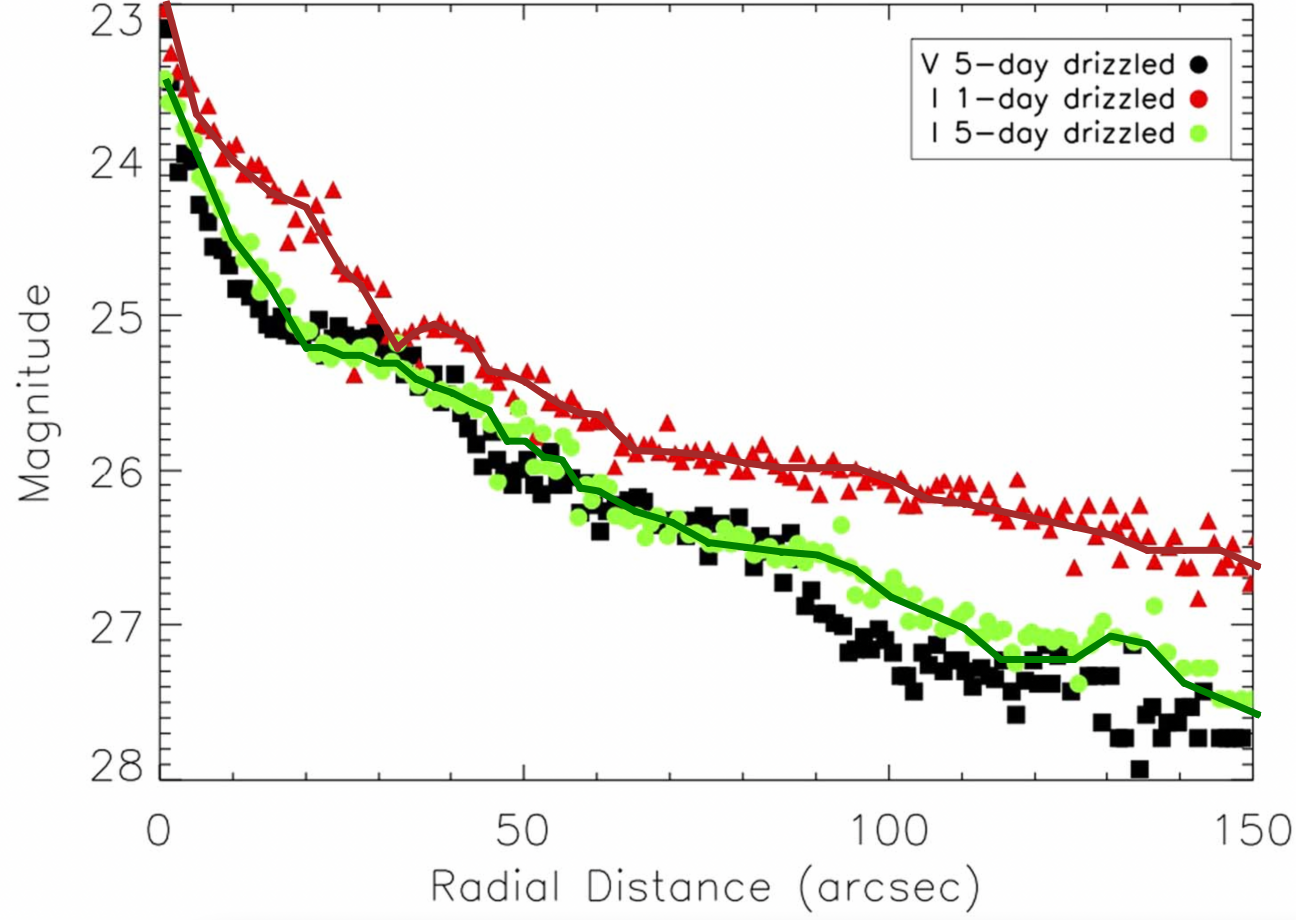}{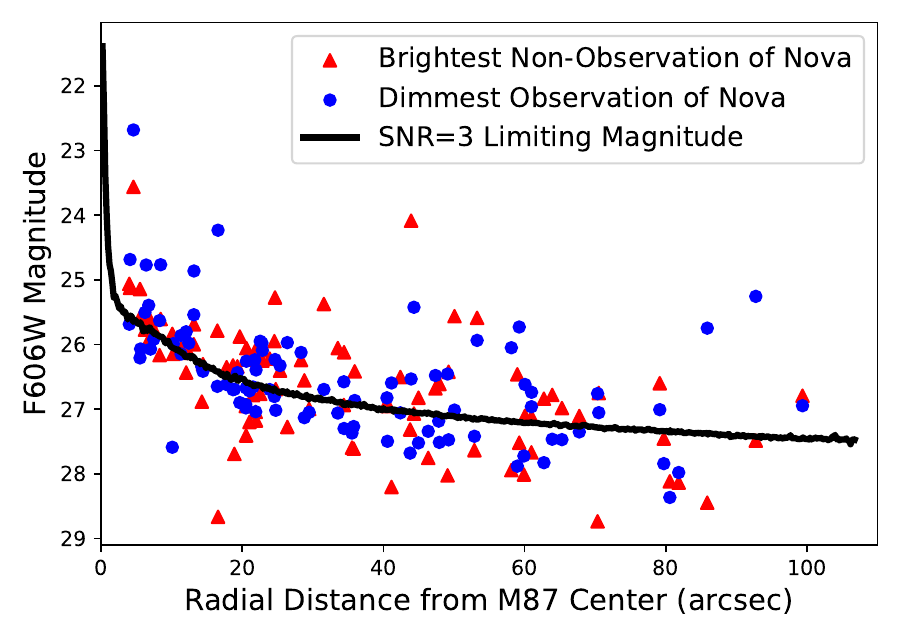}
    \caption{Left: The magnitude cutoff function used in the 2006 ACS dataset (I refers to F814W filter magnitudes while V refers to F606W filter magnitudes). The points are empirically determined 50\% recovery magnitudes from Figure 15 in \cite{Shara2016} and the overplotted lines are the magnitude cutoffs used in this study. Right: The dimmest observation, with confirmation by human eye, of a nova light-curve datapoint and brightest non-observation as a function of radius from the enter of M87. The SNR=3 line was found to bisect the overlap region between detections and non-detections well outside the inner 8" and was used as the cutoff in the 2017 WFC3 dataset in that region.}
    \label{fig:cutoffs}
\end{figure}

For each dataset, {eight} million simulated novae were placed on M87 with probability density proportional to M87's local K-band surface brightness, using the Python package \textit{emcee}'s implementation of the Metropolis-Hastings algorithm and the Large Galaxy Atlas's elliptical isophote analysis of M87 K-band light \cite{Foreman-Mackey2013, Jarrett2003}. For each simulated nova, a peak day was chosen from a one year interval containing the survey window (for the 2017 WFC3 data this year started 80 days before the first epoch, and for the 2006 ACS data this year started 200 days before the first epoch). A simulated light-curve for each nova, with a datapoint for $F606W$ and $F814W$ in each epoch of each dataset, was created by randomly choosing from the set of template light-curves used in \cite{Shara2023} and interpolating it such that its time of peak brightness aligned with the randomly chosen peak day. Shot noise was then added to each datapoint in each light-curve.

A nova was then considered detectable in a given epoch and pass band if it was brighter than a local cutoff magnitude, computed from a predetermined function of radial distance from M87's center (see Figure 3). For the 2017 WFC3 dataset, we used the same cutoff function as in \cite{Shara2023}. In the 2006 ACS dataset, we used the 50\% recovery lines found in Figure 15 of \cite{Shara2016}. Finally, simulated novae were considered undetectable if they fell within a very bright region of the jet (see Figure \ref{fig:jet-location}). These regions were chosen conservatively, containing only areas where the jet surface brightness was brighter than any background on which any real nova was successfully detected.
\newpage

The resulting simulation of the 2006 ACS dataset also yielded an implied annual nova rate of $139_{-24}^{+24}$ novae/year within the footprint of one ACS image. This coincides remarkably well with the $149_{-18.3}^{+13.4}$ novae/year finding of \cite{Shara2016}, given that different template light-curves were used in that study, and that we sampled nova coordinates proportional to K-band light instead of V-band light. This demonstrates that the current simulator's nova detection criterion is a good match to that used in \cite{Shara2016}.

\begin{figure}[h]
    \centering
    \includegraphics[height=4.1in]{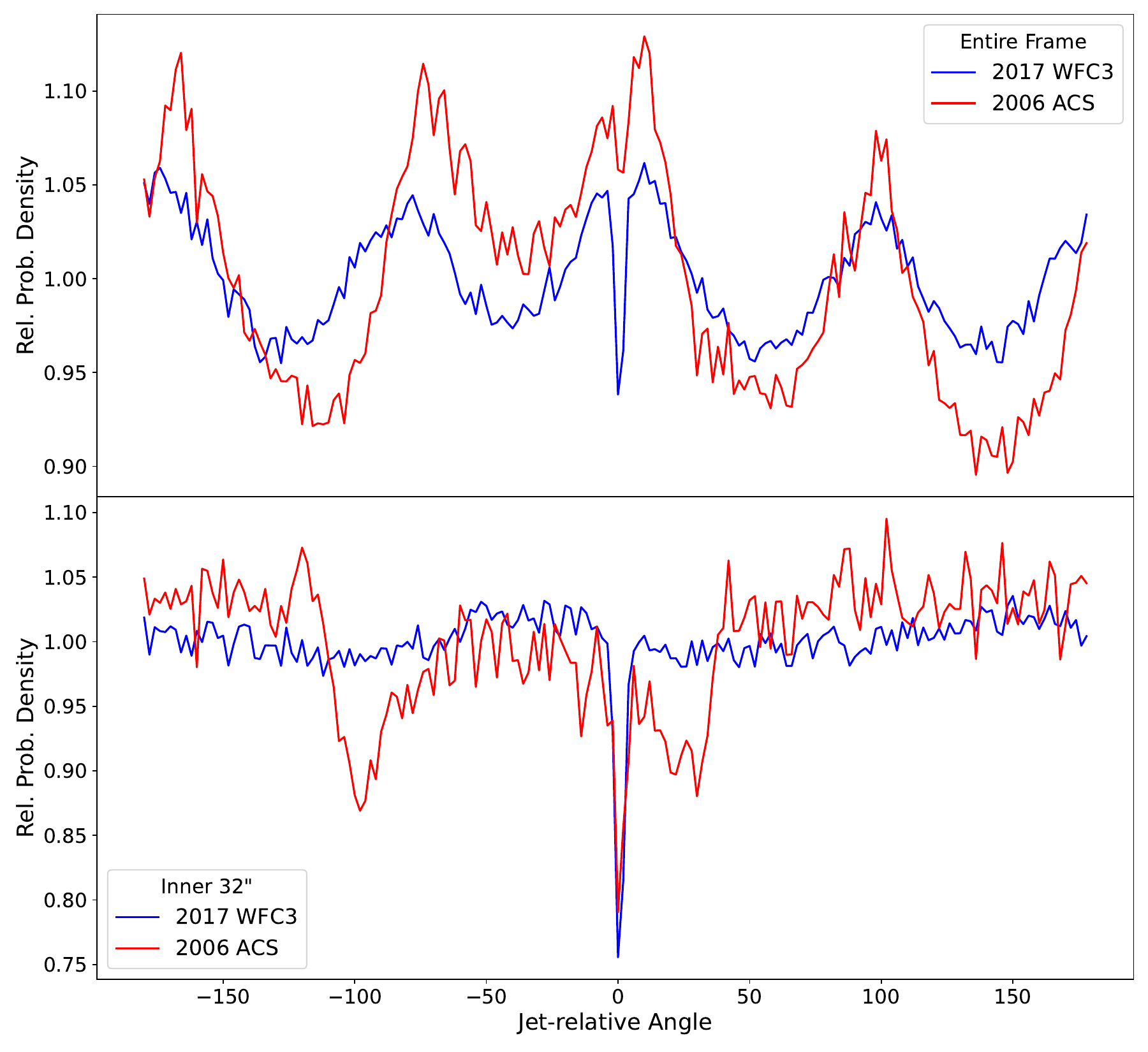}
    \caption{Variation with angle of the distribution of detected novae in simulations. Plotted is the ratio of the number of simulated novae detected per angle to the number of novae that would have been detected per angle if the distribution were uniform with angle. This is the angular probability density of detected simulated novae relative to uniform density with angle.}
    \label{fig:exp-ang-distribution}
    \vspace{-1cm}
\end{figure}    

{The output of this simulation procedure is a set of seven million ``detected" simulated novae (other ``non-detected" simulated novae were never bright enough to be detected in the images in the dataset, given their timings and locations). This sample implicitly defines an ``expected distribution" of nova detections in each dataset -- that is, the distributions that the image coordinates of successful nova detections would be drawn from if the novae ``follow the K-band light" everywhere.}

{Normalized angular marginals of these distributions are shown in Figure \ref{fig:exp-ang-distribution}.} The deviation from a uniform distribution is more pronounced in the 2006 ACS data than in the 2017 WFC3 data. This is partly because the center of the 2006 ACS images was significantly further from the center of M87 than the 2017 WFC3 dataset. In addition, the shorter duration of the 2006 survey yielded less variation in the orientation of the detector (and thus less ``smoothing") than the longer 2017 survey. Crucially, the deviation from uniformity in both surveys is never more than $\sim$ 10\%, which is much smaller than the $>$100\% deviations observed in the data shown in Figure \ref{fig:ten-wedges} and discussed in Section \ref{ssec:the-novae}.

\section{Quantifying the Observed Rate Enhancement}\label{sec:sim-rel-enhancement}

{Section \ref{sec:sim-design} defines ``expected distributions" that successful nova detections would be expected to follow under the assumption that novae ``follow the K-band light" everywhere. Does} the conclusion of section \ref{ssec:the-novae}, that the nova rate is significantly enhanced near the jet, hold up when accounting for expected deviations from radial symmetry?

We define a variety of \textit{regions of interest} (ROIs) around the jet and investigate the nova rates within them. The simulated novae generated by the procedure explained in Section 4 specify the distribution of detected novae we would expect under the assumptions that went into the simulation. We denote the fractions of detected simulated novae that fell in the ROI in each dataset as $f_{WFC3}$ and $f_{ACS}$.

\begin{figure}[h]
    \centering
    \includegraphics[width=5.3in]{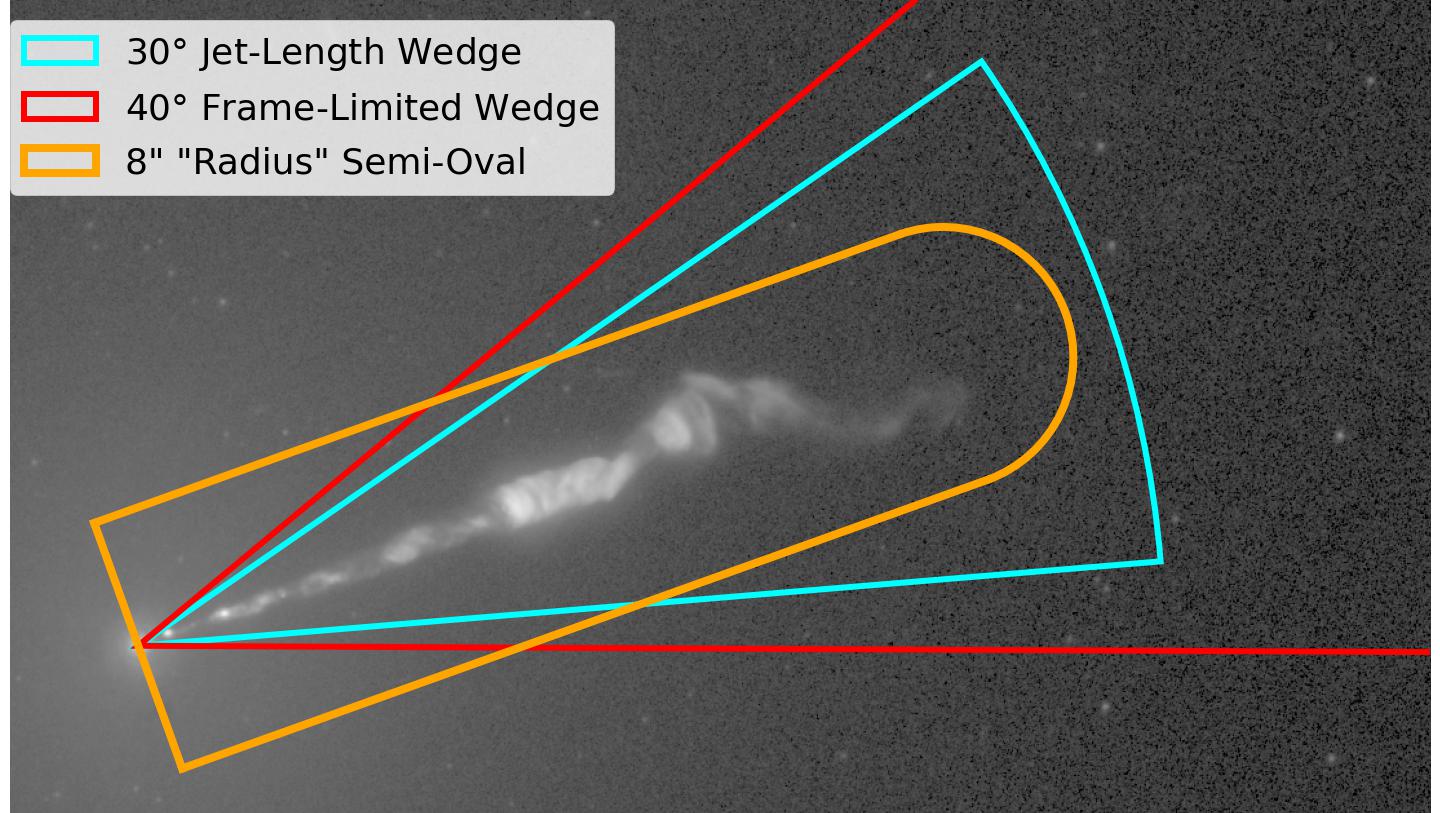}
    \caption{Shapes of the ROIs used in section \ref{sec:sim-rel-enhancement}. }
    \label{fig:shapes}
\end{figure}

\begin{figure}[h]
    \centering
    \includegraphics[width=3.4in]{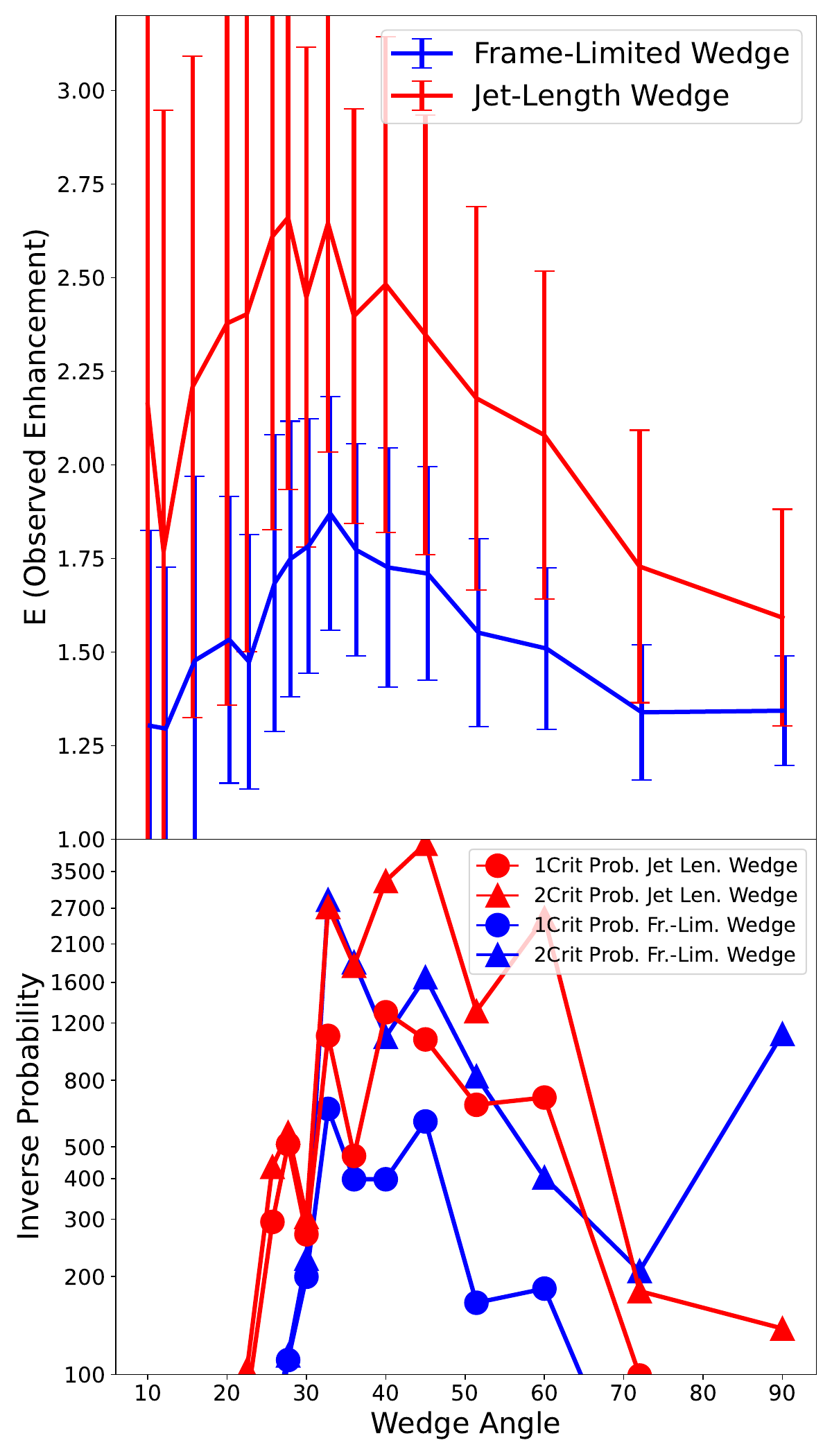}
    \includegraphics[width=3.4in]{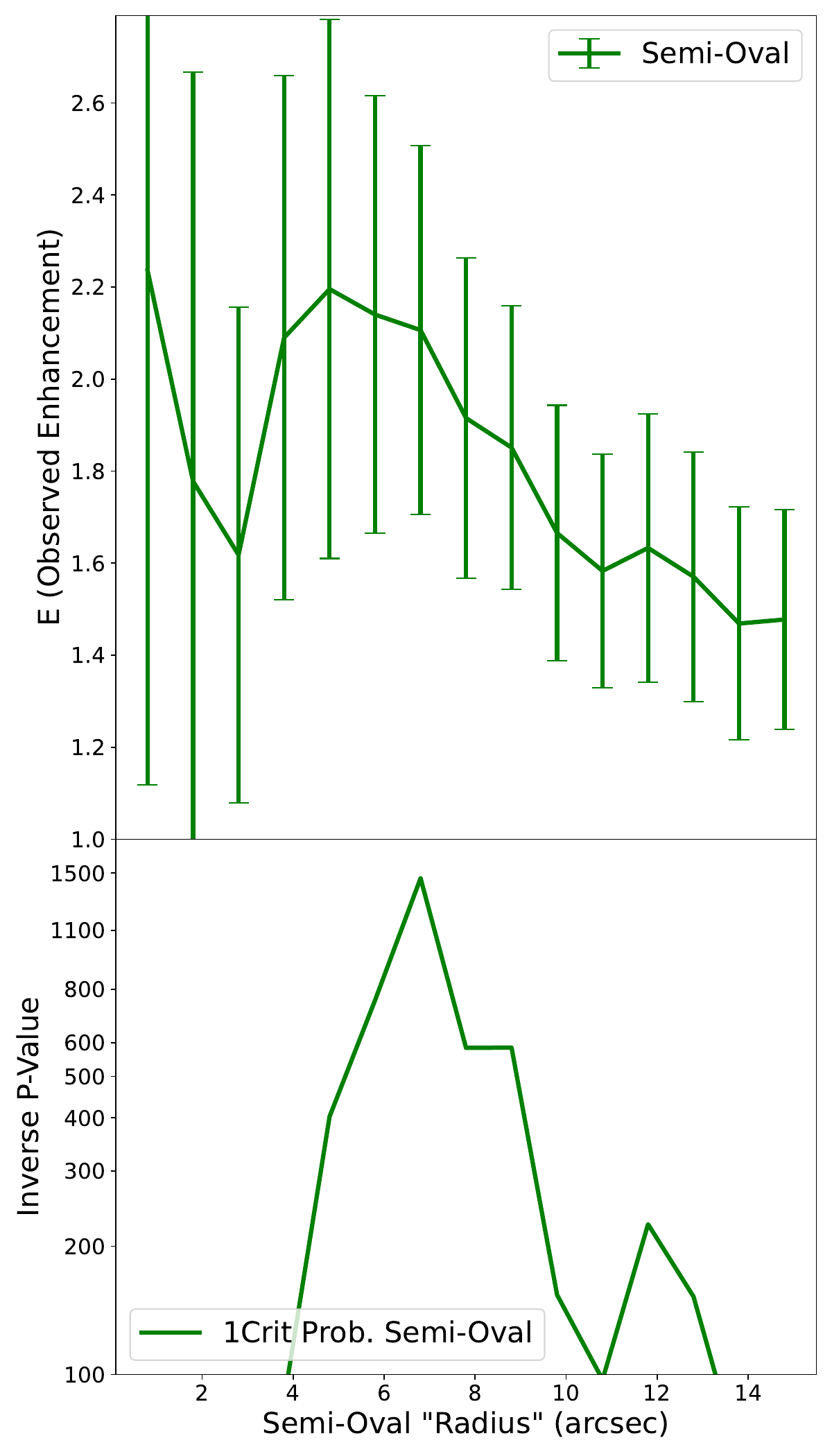}
    \caption{Top: the nova rate enhancement relative to what is expected from the distribution of detected simulated novae within wedges extending from the center of M87 out to the edge of the study footprint (blue, left) and out to 120\% of the jet length (red, left) as well as within "semi-ovals" (green, right). Bottom: the inverse of the probabilities described in section \ref{ssec:p-vals}. The shapes of the Regions of Interest used in these plots are shown in Figure \ref{fig:shapes}. {The decrease in the enhancement rate for the smallest ROIs (such as wedges less than 20 degrees wide) is included for completeness but is statistically unreliable (see discussion in section \ref{ssec:rate-enhancement-discussion}).}}
    \label{fig:enhancement-plots}
\end{figure}

We multiply these fractions by the total number of real novae detected in each dataset ($T_{WFC3} = 94$ and $T_{ACS} = 41$) to determine the number of novae we expect to have detected in the region of interest. We define the nova rate enhancement $E$ as the ratio of the actual numbers of novae observed in the region of interest ($R_{WFC3}$ and $R_{ACS}$) to the expected number (including both datasets): 
\begin{equation}\label{eqn:E-defn}
    E = \frac{R_{WFC3} + R_{ACS}}{f_{WFC3}T_{WFC3} + f_{ACS}T_{ACS}}
\end{equation}
The nova rate enhancements for a variety of wedges aligned with the jet are plotted in Figure \ref{fig:enhancement-plots}. Similarly, the rate enhancements in ``semi-ovals" -- loci of points on the same side of M87's center as the jet within a given radius of the jet center-line -- are given in Figure \ref{fig:enhancement-plots}. The greatest enhancement ratio, about 2.4, within the considered ROIs is observed in a $\sim$30 degree wide wedge extending 20\% further out than the jet (see Figure \ref{fig:shapes} to see the shape of this region).

\subsection{Statistical Significance and Uncertainty}\label{ssec:p-vals}

94 novae total were detected in the 2017 WFC3 dataset, and 41 in the 2006 ACS dataset. We thus expect that, if the datasets were recollected many times, on average, we would count $94 \, f_{WFC3}$ and $41 \, f_{ACS}$ novae in a given ROI. We thus model the numbers of novae detected in the ROI with random variables $R_{WFC3} \sim Poi(94 f_{WFC3})$ and $R_{ACS} \sim Poi(41 f_{ACS})$ (where $Poi(m)$ denotes the Poisson distribution with mean $m$). Similarly the numbers of novae detected outside the ROI are modelled as $O_{WFC3} \sim Poi\big(94 (1-f_{WFC3})\big)$ and $O_{ACS} \sim Poi\big(41 (1-f_{ACS})\big)$. The total detected nova numbers are then $T_{WFC3} = R_{WFC3} + O_{WFC3}$ and $T_{ACS} = R_{ACS} + O_{ACS}$. 

The nova rate enhancement $E$, defined in equation \ref{eqn:E-defn}, is a function of these random variables and its distribution is thus computable from their distributions. In order to quantify the statistical significance of the enhancement, we ask the question: what are the odds that we would randomly see the nova rate enhancement $E$ to be equal to or greater than the value we observed in the real data? We plot the p-value that answers this question, for each ROI, in Figure \ref{fig:enhancement-plots}. Additionally, when the ROI is one of $N$ wedges equally dividing a circle (see Figure \ref{fig:ten-wedges} to visualize this), we plot the odds that $E$ is at least the real observed value and \textit{simultaneously} the maximum enhancement seen in any other wedge is no more than its observed value (the distributions in these other wedges are modelled in the same way as the wedge containing the jet).

In order to quantify the uncertainty in the rate enhancement, we computed a ``1-sigma," 68.2 percent, bootstrap confidence interval for $E$ for each ROI. To do so, we randomly re-sampled, with replacement, 1 million sets of 94 novae from the WFC3 dataset and 41 novae from the ACS dataset. For each resulting resampled dataset, we computed $E$ as defined in equation \ref{eqn:E-defn}. This gave a distribution in $E,$ from which we computed the pivot confidence interval. These uncertainties are displayed as errorbars in Figure \ref{fig:enhancement-plots}.

\subsection{Discussion of the Rate Enhancement}\label{ssec:rate-enhancement-discussion}

The greatest nova rate enhancement is observed in wedges 120\% the length of the jet. In order to reduce noise and avoid p-hacking when choosing the size of the wedge, we average the results for wedges between 30 and 45 degrees wide. For these wedges, the average rate enhancement is $2.46_{-0.61}^{+0.61}$. The rate enhancement is highest for smaller wedges (20-30 degree widths) but is most statistically significant for wedges around 40 degrees wide, due to the number of novae included in the sample being larger. The average p-value between 30 and 45 degrees is around $\sim 1/594$ for $p_{1crit}$ and $\sim 1/1048$ for $p_{2crit}$.

{The observed nova rate enhancement decreases in very narrow wedges. In the smallest wedge we consider, a 10 degree wide jet-length wedge, the observed enhancement is $2.16_{-1.44}^{+1.44}$, lower than the enhancement of $2.46_{-0.61}^{+0.61}$ observed in 30-45 degree jet-length wedges. If the enhancement is collocated with the jet, Why should this be the case? It is possible that there is some unknown effect depressing nova rate right near the jet. However, while we mention this decrease in the enhancement for the sake of completeness, it is a statistically unreliable. The ten degree wedge is small enough that there were only 3 novae detected within it. If only one more had been detected, then the enhancement within it would have been $2.88$, higher than the enhancement observed in any wedge in the real data. In addition, as noted in section \ref{sec:sim-design}, we were deliberately conservative when estimating the impact of the jet itself in obscuring overlapping novae. Therefore, our methodology errs on the side of overestimating the density of the expected nova distribution (and thus underestimating the enhancement) in ROIs including the jet. This effect is more pronounced for the smallest ROIs, a larger fraction of which are covered by the jet.}

{It is also worth noting our analysis does not conclusively rule out an explanation for the enhancement unrelated to the jet. If we had a specific, physically motivated nova 3D spatial distribution relative to the jet, we could project this distribution into 2D coordinates and show that we \textit{could not} statistically reject this distribution whereas we \textit{could} reject some other distributions corresponding to alternative explanations for the enhancement. However, we make no specific proposal nor family of proposals for a 3D nova distribution that would correspond to a hypothesis in which the jet causes the enhancement since we have no explanation for the enhancement at this point. Additionally, we have no proposal for a distribution corresponding to a different explanation for the enhancement to be able to statically reject (see Section \ref{sec:cause}). It is possible that there is some other phenomenon unrelated to the jet causing the enhancement that happens to be aligned with the jet. While we have no reason to think this is the case, we have no data with which we can rule it out.}

\section{The Nova Distribution Outside the Jet Region}\label{sec:distr-away-from-jet}

\begin{figure}[h]
    \centering
    \includegraphics[width=5in]{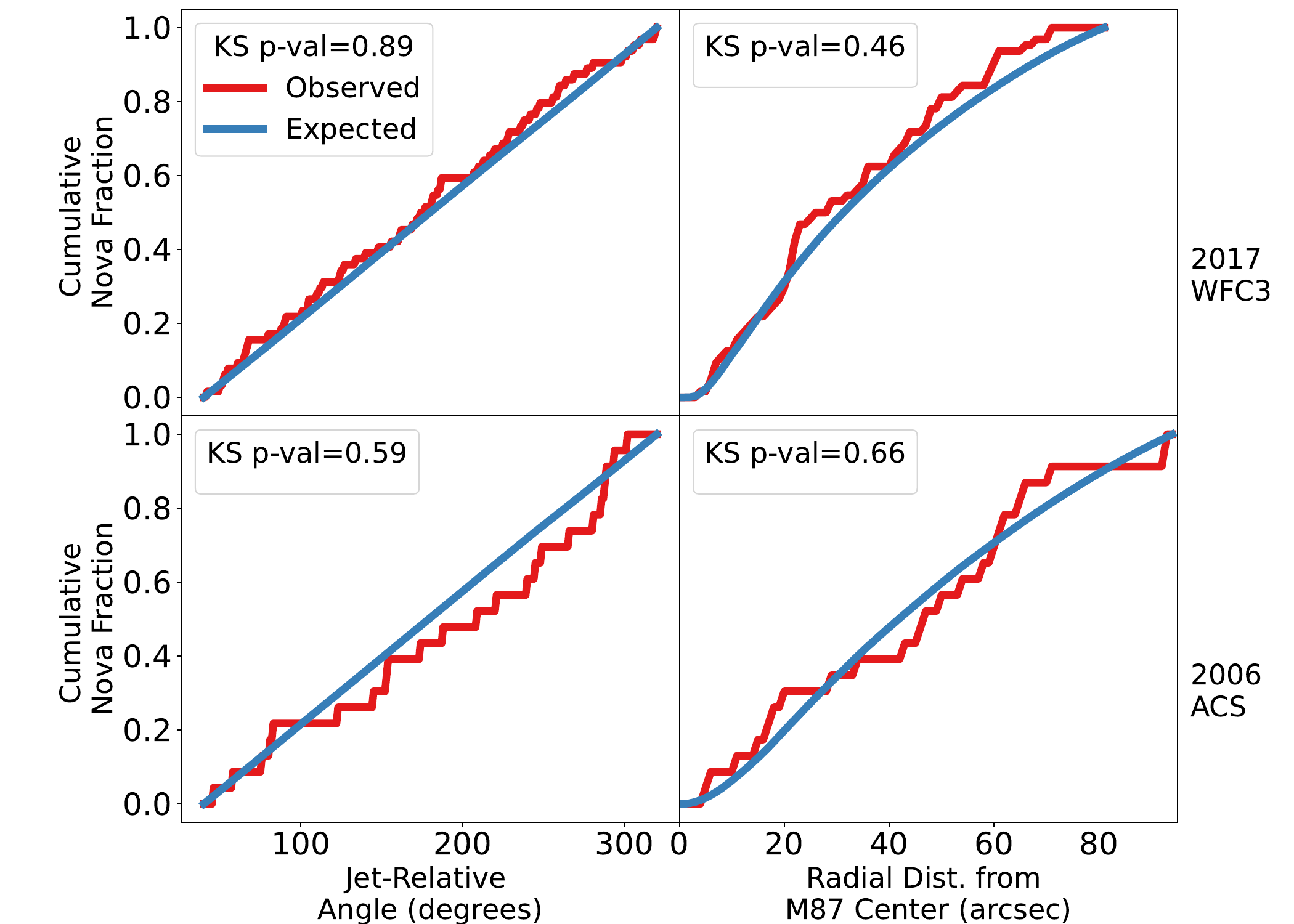}
    \caption{Cumulative angular and radial distributions of real and detected simulated novae (i.e. the expected distribution) outside of the jet region (defined here as having an angle within 40 degrees of the jet).}
    \label{fig:distr-match}
\end{figure}

The finding of a statistically significant enhancement of the nova rate near the jet relative to the expected distribution from simulation still leaves open the question: how well does the expected distribution model the nova rate in the non-enhanced region? Could the nova rate enhancement near the jet merely be part of some broader deviation from the expected distribution, on account of some unknown methodological or physical cause?

To answer this question, we compared the expected and real distributions of novae outside the region of the jet to see if any deviation of one from the other can be found. Excluding all novae with a jet-relative angle under 40 degrees (i.e. those that lie in an 80 degree frame-limited wedge), we plot the radial and angular cumulative distributions and perform KS tests. As seen in Figure \ref{fig:distr-match}, the distributions match well and no statistically significant deviation 
is observed.

\section{Differences in the Nova Population Near the Jet}\label{sec:property-differences}

\begin{figure}[h]
    \centering
    \includegraphics[width=7in]{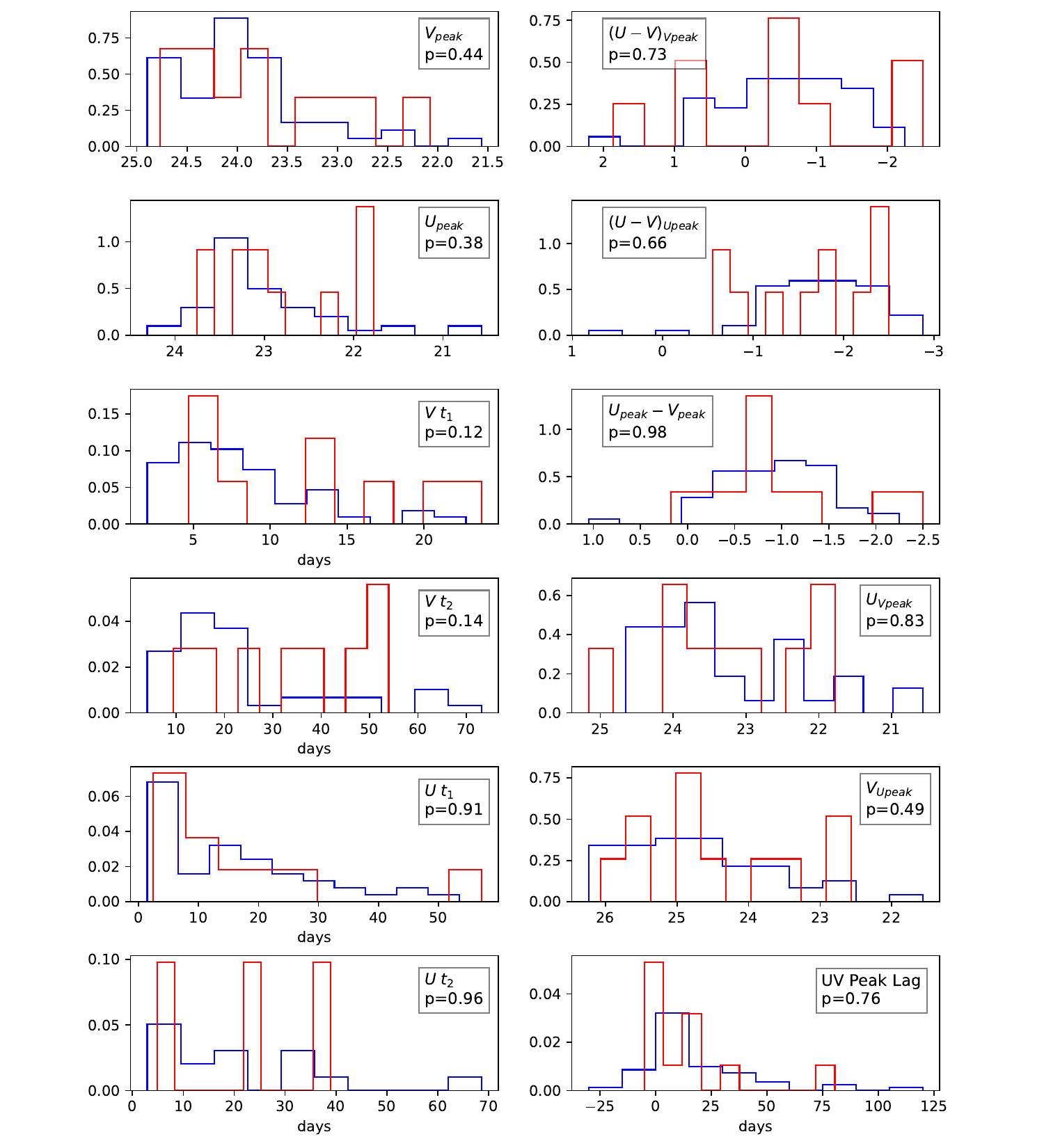}
    \caption{{Histograms of properties of novae near the jet (red) -- that is novae inside a 40-degree wedge with length 1.2 times the length of the jet (blue) -- and novae way from the jet -- outside an 80 degree infinite-length wedge. These are the same properties discussed in section \ref{sec:property-differences}. No differences in the distributions of these properties are found for these two sets of novae (p-values computed with Welch's t-test \citep{Welch1947}} are shown above).}
    \label{fig:prop-comp-hist}
\end{figure}

{The novae near the jet do not appear to have different
characteristics than those away from it.} We {compared} two sets of novae: those outside an 80 degree wide wedge centered on the jet and those within a 40 degree wide wedge. We performed Welch's t-test \citep{Welch1947} to find any statistically significant differences {in the distributions of several properties across the two sets of novae. These properties are} the peak magnitudes, color at time of peak, magnitude in one band at the time of peak brightness in the other band, $t1$ and $t2$ times, and the lag time between the peaks in different bands. We found no differences (at the $p=0.05$ level) in the novae near and away from the jet. Again, no differences were found when repeating the procedure {while taking novae within a 40 degree wide jet-length wedge as the set of novae considered ``near the jet." In Figure \ref{fig:prop-comp-hist}, these comparisons are visualized in histograms of the properties considered for the the two sets of novae.}

\section{What causes the enhanced nova rate near the M87 jet?} \label{sec:cause}

The serendipitous discovery of a type Ia supernova (SNIa) near the jet of the active galaxy 3C 78 \citep{Martel2002} prompted \citet{Livio2002} to examine whether there might be a causal connection between the jet and the stellar explosion. They noted that...`` the shock waves produced by jets might form dense clouds in a galaxy's interstellar medium (ISM), and/or that mass entrainment in the mixing layer of a jet might transport parcels of ISM to regions removed from the normal extent of the stellar populations of the galaxy." Either mechanism could conceivably enhance the accretion rate onto a massive white dwarf sufficiently to induce carbon deflagration and an SNIa event, but we are unaware of simulations or observations that support either of these scenarios. \cite{Livio2002} also noted a simpler alternative: ``Alternatively, the shock can simply heat the mass donor star and thereby increase the mass transfer rate." Again, this was not quantified.

An even simpler explanation for the concentration of novae near M87's jet might be the radiation from the jet itself. M87's jet radiates at all wavelengths. The M87 jet rivals the galaxy's nucleus in X-ray luminosity. Any Roche-lobe filling donor star in orbit about a white dwarf companion will be heated by irradiation from the jet. If that irradiation can $\sim$ double the mass transfer rate relative to that of a non-irradiated donor, then the critical hydrogen-rich envelope mass needed to trigger a classical nova thermonuclear runaway could be accreted in half the time. Novae close to the jet would then erupt $\sim$ twice as often as novae elsewhere in M87... as is observed in our analysis. 

Unfortunately this appealingly simple explanation does not work. The luminosity of M87's jet is of order $10^{42}$ erg/s (i.e. $10^{9} L_{\odot} ~$\citep{Punsley2023}). To substantially increase the mass transfer rate in a cataclysmic binary, the donor must be ``swollen" by an atmospheric scale height or more \citep{Livio1987,Kovetz1988,Ritter1988,Hillman2020}, which requires the input of at least $\sim 0.1 - 0.01 L_{\odot} $. The cross section of a typical cataclysmic binary red dwarf (of order [$R_{\odot}/3]^{2}$ $\sim$ $10^{21} ~cm^{2}$) means that even if all of the jet's luminosity was emitted by a single point, a cataclysmic binary would have to be closer than  $\sim$ $10^{-2}$ pc to receive enough irradiation to enhance the mass transfer rate enough to significantly shorten the time between nova eruptions. 

Another possibility is that stars (including cataclysmic binaries) are forming under the influence of a jet \citep{DeYoung1989,Klamer2004,Gaibler2012,Duncan2023}. While such stars would migrate away in all directions, their orbits would tend to return them to the neighborhoods of their birth - the jet - a few times per Gyr. The space density of novae would thus be maximized in M87 near its jet. This appealingly simple explanation fails to account for the lack of enhancement of novae in the direction of M87's counterjet \citep{Sparks1992}.

{A final possibility is that M87 has ``recently" accreted a satellite along the direction of the jet, leading to a surplus of multiple populations, including novae, along the orbit of the incoming satellite. Indeed, \citet{Arnaboldi2016} have written that ``A substructure detected in the projected phase-space of the line-of-sight velocity vs. major axis distance for the M87 halo planetary nebulae provides evidence for the recent accretion event of a satellite galaxy with luminosity twice that of M33. The satellite stars were tidally stripped about 1 Gyr ago, and reached apocenter at a major axis distance of 60-90 kpc from the center of M87." The dynamical crossing time of M87 (which is $\sim$ 40 kpc in extent at its 25th magnitude B-band isophote) is of order 1 Gyr, so one could imagine that a concentration of novae created by such an accretion event might not yet have been dynamically disrupted. The nova rate of a galaxy with twice M33's luminosity is $\sim$ 5/yr \citep{Williams2004}, so during our 9 monthlong observing campaign of 40\% of M87's light, one might expect an enhancement of just $\sim$ 5 x (9/12) x 0.4 = 1.5 novae along the orbit of the disrupted satellite, an order of magnitude too small to be a plausible explanation of the observed jet nova enhancement.

While it is possible that there is no causal connection between the jet and the ``excess" novae that we observe, the remarkable spatial coincidence between them is established. We conclude that the enhanced rate of novae along M87's jet is almost certainly real, and it remains unexplained.} 

\section{Conclusions}

A map of the locations of the 135 novae we have discovered in our {\it HST} surveys of M87 reveals a striking concentration of novae near that galaxy's iconic jet, and no enhancement in the direction of its counterjet. Detailed simulations to account for the non-sphericity of M87, and the slightly variable placement of M87's nucleus on the the telescope's detectors, confirm the results of 
simple analytic and numerical simulations
: the distribution of classical novae in M87 is strongly concentrated towards the galaxy's jet. The probability of a random occurrence of the observed distribution varies with the shapes and sizes of sub-regions selected, but is typically in the range of 0.1\% to 1\%. 

The suggestion that irradiation by the jet of the hydrogen-rich donor stars in cataclysmic binaries drives enhanced mass transfer fails by many orders of magnitude. The hypothesis of enhanced star formation (including cataclysmic binaries) triggered by the jet is appealingly simple, and could enhance the space density of novae near the jet of M87. The same hypothesis also suggests an enhancement of novae near M87's counterjet, which is not observed. The enhanced rate of novae along M87's jet is now firmly established, and unexplained.

\section{Acknowledgements}
This research is based on observations made with the NASA/ESA Hubble Space Telescope obtained from the Space Telescope Science Institute, which is operated by the Association of Universities for Research in Astronomy, Inc., under NASA contract NAS 5–26555.  These observations are associated with programs 10543 (PI:Baltz) and 14618 (PI:Shara). The specific observations analyzed can be accessed via \dataset[https://doi.org/10.17909/natk-ks60]{https://doi.org/10.17909/natk-ks60}.  Support to MAST for these data is provided by the NASA Office of Space Science via grant NAG5–7584 and by other grants and contracts. MMS and RH were funded by NASA/STScI grant GO-14651. The paper also is based upon work of RH supported by NASA under award number 80GSFC21M0002.

This work has made use of data from the European Space Agency (ESA) mission
{\it Gaia} (\url{https://www.cosmos.esa.int/gaia}), processed by the {\it Gaia}
Data Processing and Analysis Consortium (DPAC,
\url{https://www.cosmos.esa.int/web/gaia/dpac/consortium}). Funding for the DPAC
has been provided by national institutions, in particular the institutions
participating in the {\it Gaia} Multilateral Agreement.

\bibliography{M87_distribution}{}
\bibliographystyle{aasjournal}

\section{Appendix}

\begin{deluxetable}{ccccc}\label{tab:updated-pos}
\tablecaption{\edit1{Updated positions of the novae in Table 3 of \citet{Shara2016}.}}
\tablehead{\colhead{Nova ID} & \colhead{Updated R.A.} & \colhead{Updated Decl.} & \colhead{Old R.A.} & \colhead{Old Decl.} \\ \#  & HH:MM:SS.ss & DD:MM:SS.ss & HH:MM:SS.ss & DD:MM:SS.ss}
\startdata
1 & 12:30:48.84 & +12:23:15.92 & 12:30:48.85 & +12:23:16.91 \\
2 & 12:30:49.84 & +12:22:55.27 & 12:30:49.85 & +12:22:56.80 \\
3 & 12:30:54.20 & +12:22:02.09 & 12:30:54.23 & +12:22:03.05 \\
4 & 12:30:49.59 & +12:23:23.73 & 12:30:49.60 & +12:23:24.82 \\
5 & 12:30:46.85 & +12:23:46.98 & 12:30:46.86 & +12:23:47.97 \\
6 & 12:30:50.37 & +12:23:24.76 & 12:30:50.38 & +12:23:25.86 \\
7 & 12:30:49.91 & +12:23:19.99 & 12:30:49.92 & +12:23:20.93 \\
8 & 12:30:48.16 & +12:23:30.41 & 12:30:48.18 & +12:23:31.37 \\
9 & 12:30:49.15 & +12:23:45.30 & 12:30:49.16 & +12:23:46.32 \\
10 & 12:30:49.66 & +12:23:44.44 & 12:30:49.68 & +12:23:45.43 \\
11 & 12:30:51.93 & +12:23:56.33 & 12:30:51.94 & +12:23:57.37 \\
12 & 12:30:50.54 & +12:24:58.72 & 12:30:50.54 & +12:24:59.69 \\
13 & 12:30:46.13 & +12:22:36.80 & 12:30:46.14 & +12:22:37.80 \\
14 & 12:30:48.47 & +12:23:36.67 & 12:30:48.48 & +12:23:37.68 \\
15 & 12:30:49.46 & +12:23:22.73 & 12:30:49.47 & +12:23:23.77 \\
16 & 12:30:53.17 & +12:23:43.88 & 12:30:53.18 & +12:23:44.84 \\
17 & 12:30:43.79 & +12:23:33.66 & 12:30:43.79 & +12:23:34.65 \\
18 & 12:30:47.72 & +12:24:22.56 & 12:30:47.73 & +12:24:23.67 \\
19 & 12:30:49.07 & +12:23:27.84 & 12:30:49.07 & +12:23:28.82 \\
20 & 12:30:51.37 & +12:23:32.07 & 12:30:51.38 & +12:23:33.04 \\
21 & 12:30:47.55 & +12:22:51.77 & 12:30:47.56 & +12:22:52.73 \\
22 & 12:30:46.21 & +12:23:20.91 & 12:30:46.23 & +12:23:21.99 \\
23 & 12:30:47.88 & +12:23:54.01 & 12:30:47.89 & +12:22:54.95 \\
24 & 12:30:49.47 & +12:23:08.79 & 12:30:49.48 & +12:23:09.76 \\
25 & 12:30:47.07 & +12:24:10.68 & 12:30:47.09 & +12:24:11.68 \\
26 & 12:30:47.75 & +12:21:59.15 & 12:30:47.76 & +12:22:00.15 \\
27 & 12:30:44.26 & +12:24:25.13 & 12:30:44.25 & +12:24:26.28 \\
28 & 12:30:49.73 & +12:24:21.12 & 12:30:49.73 & +12:24:22.06 \\
29 & 12:30:54.51 & +12:21:37.85 & 12:30:54.53 & +12:21:38.70 \\
30 & 12:30:49.61 & +12:25:05.88 & 12:30:49.62 & +12:25:06.81 \\
31 & 12:30:46.78 & +12:22:57.01 & 12:30:46.79 & +12:22:58.00 \\
32 & 12:30:46.92 & +12:22:24.20 & 12:30:46.93 & +12:22:35.26 \\
33 & 12:30:45.03 & +12:23:09.05 & 12:30:45.04 & +12:23:10.08 \\
34 & 12:30:45.30 & +12:23:43.61 & 12:30:45.31 & +12:23:44.69 \\
35 & 12:30:53.17 & +12:23:00.01 & 12:30:53.18 & +12:23:01.07 \\
36 & 12:30:44.36 & +12:23:01.01 & 12:30:44.36 & +12:23:02:03 \\
37 & 12:30:53.45 & +12:23:35.18 & 12:30:53.46 & +12:23:36.23 \\
38 & 12:30:45.47 & +12:23:48.08 & 12:30:45.47 & +12:24:49.12 \\
39 & 12:30:44.93 & +12:23:54.12 & 12:30:44.94 & +12:23:55.18 \\
40 & 12:30:45.89 & +12:23:33.17 & 12:30:45.90 & +12:23:34.29 \\
41 & 12:30:46.68 & +12:22:37.52 & 12:30:46.68 & +12:22:38.69
\enddata
\end{deluxetable}

\end{document}